\documentclass[prb,aps,twocolumn,draft,tightenlines,
showpacs,floatfix,superscriptaddress]{revtex4}
\usepackage{psfig}
\usepackage{amssymb}
\usepackage{amsmath}
\usepackage{colordvi}

\begin{document}

\title{Control of spin relaxation in semiconductor double quantum dots}
\author{Y. Y. Wang}
\affiliation{Hefei National Laboratory for Physical Sciences at
  Microscale, University of Science and Technology of China, Hefei,
  Anhui, 230026, China}
\affiliation{Department of Physics,
University of Science and Technology of China, Hefei,
  Anhui, 230026, China}
\altaffiliation{Mailing Address}
\author{M. W. Wu}
\thanks{Author to whom correspondence should be addressed}%
\email{mwwu@ustc.edu.cn.}
\affiliation{Hefei National Laboratory for Physical Sciences at
  Microscale, University of Science and Technology of China, Hefei,
  Anhui, 230026, China}
\affiliation{Department of Physics,
University of Science and Technology of China, Hefei,
  Anhui, 230026, China}
\altaffiliation{Mailing Address}

\date{\today}
\begin{abstract}
We propose a scheme to manipulate the spin relaxation in vertically
coupled semiconductor double quantum dots. Up to {\em twelve} orders
of magnitude variation  of the spin relaxation time can be achieved
by a small gate voltage applied vertically on the double dot.
Different effects such as the  dot size,
barrier height, inter-dot distance, and magnetic field on
the  spin relaxation are investigated in detail. The 
condition to achieve a large variation is discussed.

\end{abstract}

\maketitle

Spin related phenomena in semiconductor nanostructures have attracted
much interest
recently due to the fast growing field of spintronics \cite{Awschalom}.
Among different structures, quantum dots (QDs) have caused
a lot of attention
as they provide a versatile system to manipulate the spin and/or electronic
states \cite{Hans}. Many proposals of spin qubits, spin filters,
spin pumps and spin quantum
gates are proposed and/or demonstrated based on different kinds of QDs
\cite{Hans,Barenco,Loss,Burkard,das, Folk,Aono, Recher, Ernesto, Romo}.
Manipulation and understanding of the spin coherence in QDs
are of great importance in the design and
the operation of these spin devices. There are many theoretical and 
experimental investigations on the spin relaxation in single 
QDs \cite{Alexander, Governale, Woods, Cheng,Tsitsishvili,Golovach,hanson},
 double QDs \cite{johnson,Peter} and quasi-one-dimension coupled
QDs \cite{Tamborenea} due to
the Dresselhaus or Rashba spin-orbit couplings \cite{Dresselhaus, Yu}.
In this paper, we propose a feasible and convenient way to
manipulate the spin relaxation
in double QDs by a small gate voltage. 
We show that up to {\em twelve} orders of magnitude variation of the
longitudinal spin relaxation time (SRT) can be tuned in
such a system.

\begin{figure}[htb]
 \centerline
  {\psfig{figure=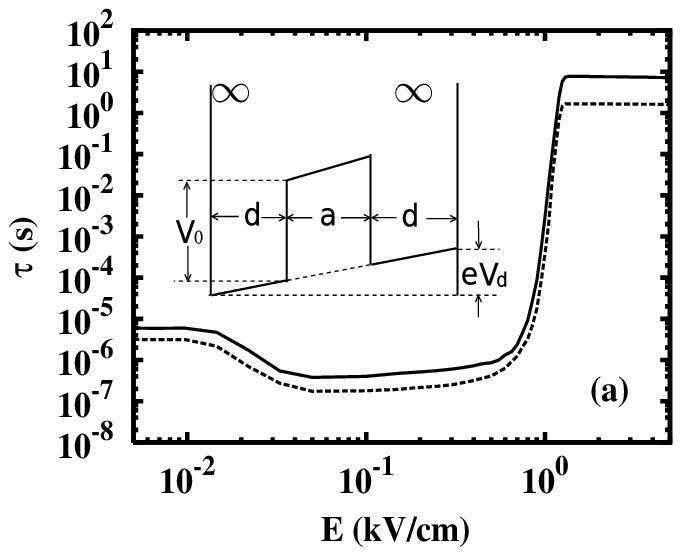,width=6.4cm,height=5.2cm,angle=0}}
\vskip -0.25cm
\centerline
  {\psfig{figure=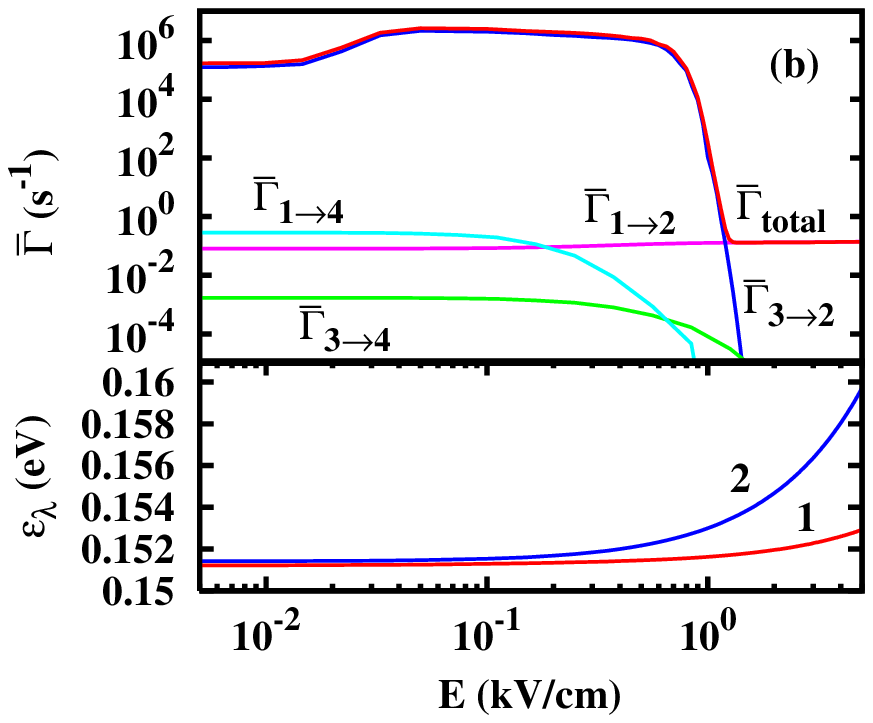,width=6.4cm,height=5.3cm,angle=0}}
\caption{(Color online) (a) SRT {\em vs.} the electric field. Solid
 curve: perturbation result; Dotted curve: exact
   diagonalization result;
Inset: Schematic of the potential along the vertical ($z$) direction.
(b) Upper panel:
 Weighted scattering rates $\bar\Gamma_{i\to j}$
 between different energy levels (from ``spin-up''
 to ``spin-down'') {\em vs.} the electric field. $\bar\Gamma_{\mbox{total}}$ is the
total weighted scattering rate from the ``spin-up'' to the
''spin-down'' states.
  Lower panel: Energy level $\varepsilon_{\lambda}$ of the $z$ direction
of the double QD {\em vs.} the electric field. }
\label{fig1}
\end{figure}

We consider a single electron spin in two vertically coupled QDs.
Each QD is  confined by a parabolic potential $V_{c}({\bf r})
=\frac{1}{2} m^{\ast} \omega_{0}^{2} {\bf r}^{2}$
(Therefore the effective dot diameter
$d_{0}=\sqrt{\hbar\pi/m^{\ast} \omega_{0}}$) along the
$x$-$y$ plane in a quantum well of width $d$ with its growth direction
along the $z$-axis. A gate voltage $V_d$ together with a magnetic field $B$
are applied along the growth direction.
A schematic of the potential of the coupled quantum wells is
plotted in the inset of Fig.\ 1(a) and the potential is given by\cite{comment}
\begin{equation}
V_{z}(z)=
\begin{cases} eEz+\frac{1}{2}eV_{d}, & \frac{1}{2}a<|z|<\frac{1}{2}a+d\\
eEz+\frac{1}{2}eV_{d} +V_{0}, &|z|\leqslant\frac{1}{2}a \\
\infty, & \mbox{otherwise}
\end{cases}
\end{equation}
in which $V_{0}$ represents the barrier height between the
two coupled QDs, $a$ is the
barrier width and $E=V_{d}/(a+2d)$ denotes the electric field due to
the gate voltage.
The origin of the $z$-axis is chosen to be the center of the barrier
between the two QDs. By solving the Schr\"odinger equations along
the $z$-axis
${d^{2}\psi_{z}}/{d \xi_{i}^{2}}-\xi_{i} \psi_{z} =0$
with $\xi_{1}=2^{1/3}(\frac{m^{\ast}}{\hbar^{2}
  e^2E^{2}})^{1/3}(eEz-\varepsilon+eV_{d}/2)$ for
$\frac{1}{2}a<|z|<\frac{1}{2}a+d$
and $\xi_{2}=2^{1/3}(\frac{m^{\ast}}{\hbar^{2}
  e^2E^{2}})^{1/3}(eEz-\varepsilon+eV_{d}/2+V_{0})$ for
$|z|\leqslant\frac{1}{2}a$,
one obtains the wave function:
\begin{equation}
\label{psiz}
\psi_{z}(z)=
\begin{cases}
A_{1} \mbox{Ai}(\xi_{1})+A_{2}\mbox{Bi}(\xi_{1}),&
-(\frac{a}{2}+d)<z<-\frac{a}{2}\\
B_{1} \mbox{Ai}(\xi_{2})+B_{2}\mbox{Bi}(\xi_{2}),&
|z|\leqslant\frac{1}{2}a\\
C_{1} \mbox{Ai}(\xi_{1})+C_{2}\mbox{Bi}(\xi_{1}),&
\frac{a}{2}<z<\frac{a}{2}+d\\
\end{cases}
\end{equation}
in which Ai and Bi are the Airy functions. The coefficients together with the
eigenenergy $\varepsilon_\lambda$ can be obtained from the boundary
conditions $\psi_{z\lambda}(z=\pm(a/2+d))=0$, the continuity conditions
at $z=\pm \frac{1}{2}a$ and the condition of normalization $\int
\psi_{z\lambda}^\ast(z) \psi_{z\lambda}(z) dz =1$.
The electron Hamiltonian in the $x$-$y$ plane is $H_{e}=H_{0}+H_{so}$, where
$H_{0}=(P_x^{2}+P_y^2)/(2m^\ast)+V_c({\bf r})+H_B$ is electron
Hamiltonian without the spin-orbit interaction, in which ${\bf
  P}\equiv(P_x,P_y)
  = -i \hbar \nabla + (e/c)\bf{A}$ with ${\bf A} =
  (B/2)(-y,x)$ is the electron momentum operator. $m^{\ast}$
  is the electron effective mass.
$H_{B}=\frac{1}{2}g\mu_{B}B\sigma_{z}$ is the Zeeman energy with
$\sigma_z$ denoting the Pauli matrix.
 $H_{so}=\frac{\gamma}{\hbar^{3}}\sum_\lambda
\langle P_{z}^{2} \rangle_\lambda (-P_{x}\sigma_{x}+P_{y}\sigma_{y})$ is the
Dresselhaus spin-orbit coupling \cite{Dresselhaus} with $\langle P_{z}^{2} \rangle_\lambda
\equiv -\hbar^2\int \psi_{z\lambda}^\ast(z)\partial^2/\partial z^2\psi_{z\lambda}(z)dz$ and
$\gamma=27.5$ \AA$^{3}\cdot$eV \cite{Knap}.
For the small applied gate voltage, the
Rashba spin-orbit coupling \cite{Yu} is unimportant in this study \cite{flat}.
The eigenenergy of $H_{0}$ is
$E_{nl\sigma} = \hbar \Omega (2n+|l|+1)-\hbar l \omega_{B}+\sigma
E_{B}$, in which $\Omega = \sqrt{\omega_{0}^{2}+\omega_{B}^{2}}$,
$\omega_{B}=eB/(2m^{\ast})$ and $E_{B}= \frac{1}{2} g \mu_{B} B $.
The eigenfunction $\langle{\bf r}|nl\sigma\rangle = N_{n,l} (\alpha
r)^{|l|} e^{-(\alpha r)^{2}}
L_{n}^{|l|} ((\alpha r)^{2}) e^{il\theta} \chi_{\sigma} $ with $N_{n,l}=(\alpha^{2}
n!/\pi(n+|l|)!)^{1/2}$ and $\alpha=\sqrt{m^{\ast}
\Omega/\hbar}$. $L_{n}^{|l|}$ is the generalized Laguerre
polynomial. $\chi_{\sigma}$ represents the eigenfunction of
$\sigma_{z}$.
In these equations $n=0,1,2,\cdots$, $l=0,\pm1,\pm2,\cdots$ and
$\sigma =\pm 1$ are quantum numbers.
From the eigenfunction of $H_{0}$, one can construct the wave
function $|\Psi_{\ell}\rangle$ of $H_{e}$ by either the perturbation
calculations \cite{Alexander,Woods} modified by the right energy
corrections pointed out by Cheng {\em et al.} \cite{Cheng} or the exact
diagonalization approach.\cite{Cheng}

The SRT $\tau$ is calculated from $\tau^{-1}=\sum_{if} f_{i} \Gamma_{i \to f}$
in which $f_{i}=C \exp[-E_{i}/(k_{B}T)]$ denotes the Maxwell
distribution of the $i$-th level with $C$ standing for the normalization parameter
and
\begin{eqnarray}
\Gamma_{i \to f}&=& \frac{2\pi}{\hbar} \sum_{{\bf q}
 {\lambda_{1}}}  |M_{{\bf q} {\lambda_{1}}}|^{2}
|\langle f|e^{i{\bf q}\cdot {\bf r}}|i \rangle |^{2}
\Big[\bar{n}_{{\bf q} {\lambda_{1}}} \delta(E_{f}-E_{i}\nonumber\\
&&\hspace{-0.3cm}\mbox{}-\hbar \omega_{{\bf q} {\lambda_{1}}}) +
  (\bar{n}_{{\bf q} {\lambda_{1}}}+1)
\delta(E_{f}-E_{i}+\hbar \omega_{{\bf q} {\lambda_{1}}})\Big]
\label{rate}
\end{eqnarray}
is the transition rate from the $i$-th level to the $f$-th one due to the
electron-phonon scattering due to the deformation potential  with
$|M_{{\bf q}sl}|^{2} = \hbar \Xi^{2}q/2Dv_{sl}$ and  the piezoelectric coupling
for the longitudinal  phonon mode with $|M_{{\bf q}
  pl}|^{2}= (32\hbar \pi^{2}
e^{2}e_{14}^{2}/\kappa^{2} D v_{sl})[(3q_{x}q_{y}q_{z})^{2}/q^{7}]$
and for the two transverse  phonon modes with $\sum_{j=1,2}|M_{{\bf
    q}pt_j}|^{2} =(32\hbar \pi^{2}
e^{2}e_{14}^{2}/\kappa^{2} D
v_{st}q^{5})[q_{x}^{2}q_{y}^{2}+q_{y}^{2}q_{z}^{2} +
q_{z}^{2}q_{x}^{2} -(3q_{x}q_{y}q_{z})^{2}/q^{2}]$.
$\bar{n}_{{\bf q}{\lambda_{1}}}$ represents the Bose distribution of phonon
with mode ${\lambda_{1}}$ and momentum ${\bf q}$ at the temperature $T$.
Here $\Xi=7$\ eV stands for the acoustic deformation potential;
$D=5.3 \times 10^{3}$\ kg/m$^{3}$ is the GaAs volume
density; $e_{14}=1.41 \times 10^{9}$\ V/m
 is the piezoelectric constant and $\kappa=12.9$ denotes
the static dielectric constant. The acoustic phonon spectra are given
by $\omega_{{\bf q}ql} = v_{sl} q$ for the longitudinal mode and
$\omega_{{\bf q}pt} = v_{st} q$ for the transverse modes with
$v_{sl}=5.29 \times 10^{3}$\ m/s and $v_{st}=2.48 \times 10^{3}$\ m/s
being the corresponding sound velocities.

The states $i$ and $f$ in Eq.\ (\ref{rate}) are the eigenstates of the
Hamiltonian $H_e$. In order to demonstrate the  physics clearly, we
first use the corrected perturbation method by Cheng {\em et al.} \cite{Cheng}
to study the SRT. For the double dot system, we
need to include the lowest two energy
levels of z direction which we label as $|1_z\rangle$ and $|2_z \rangle$
[Eq.\ (\ref{psiz})].
In $x$-$y$ plane, the lowest six energy
levels of $H_{0}$ for each QD are considered, {\em i.e.}, $|00+\rangle$,
$|00- \rangle$,
$|01+\rangle$, $|01-\rangle$, $|0-1+\rangle$, and
$|0-1-\rangle$.
The wave functions of the lowest four states of
$H_{e}+\frac{P_{z}^{2}}{2 m^{\ast}}+V_{z}$
constructed from these levels  are
therefore given by
\begin{eqnarray}
|\Psi_{1}\rangle&=&|00+\rangle|1_z\rangle
-{\cal B}_1|0-1-\rangle|1_z \rangle\ ,\\
|\Psi_{2}\rangle&=&|00-\rangle|1_z\rangle -{\cal A}_1|01+
\rangle|1_z \rangle\ ,\\
|\Psi_{3}\rangle&=&|00+\rangle|2_z\rangle
-{\cal B}_2|0-1-\rangle|2_z \rangle \,\\
|\Psi_{4}\rangle&=&|00-\rangle|2_z\rangle
 -{\cal A}_2|01+\rangle|2_z \rangle \ ,
\end{eqnarray}
with the corresponding energies being:
\begin{eqnarray}
\label{E1}
E_{1}&=&E_{00+,1}-|{\cal B}_1|^{2}(E_{0-1-,1}-
E_{00+,1})\ ,\\
\label{E2}
E_{2}&=&E_{00-,1}-|{\cal A}_1|^{2}(E_{01+,1}
-E_{00-,1})\ ,\\
\label{E3}
E_{3}&=&E_{00+,2}-|{\cal B}_2|^{2}(E_{0-1-,2}
-E_{00+,2})\ ,\\
\label{E4}
E_{4}&=&E_{00-,2}-|{\cal
  A}_{2}|^{2}(E_{01+,2}-E_{0-,2})\ .
\end{eqnarray}
In these equations $E_{nl\sigma,\lambda}=E_{nl\sigma}+\varepsilon_\lambda$;
${\cal B}_{\lambda}=i\alpha \gamma_{\lambda}^{\ast}
(1-eB/(2\hbar
\alpha^{2}))/ (E_{0-1-,\lambda}-E_{00+,\lambda})$ and
${\cal A}_{\lambda}=i\alpha \gamma_{\lambda}^{\ast}(1+eB/(2\hbar
\alpha^{2}))/(E_{01+,\lambda}-E_{00-,\lambda})$
with $\gamma_{\lambda}^{\ast}=\gamma \langle
P_{z}^{2}\rangle_{\lambda}/\hbar^2$.  $\lambda$($=1,2$) is the quantum
number of $z$-axis.
Now we calculate the spin-flip rates
from the ``spin-up'' states $|\Psi_{2m-1}\rangle$ to the ``spin-down''
ones $|\Psi_{2m}\rangle$ ($m=1,2$) due to the electron-phonon scattering.
There are nine spin-flip scattering rates.
The scatting rate from the ``spin-up''  state $i$ to the ``spin-down'' one
$f$ reads
\begin{eqnarray}
\Gamma_{i \to f}&=& |{\cal A}_{f}-{\cal B}_{i}|^{2}
\{n_{q}+[1+ \mbox{sgn}(i-f)]/2\}
 q^{3}\nonumber\int_{0}^{\pi/2}d\theta \\
&&\hspace{-1cm} \mbox{}
\times \Big[C_{LD}q^{2}\sin^{3}\theta+ C_{LP}q^{2}\sin^{7}\theta \cos^{2}
\theta + C_{TP} \sin^{5}\theta\nonumber\\
&&\hspace{-1cm} \mbox{}
\times (\sin^{4}\theta+8\cos^{4}\theta)\Big]
e^{-q^{2}\sin^{2}\theta/2} |I_{if}(q \cos\theta)|^{2}\ ,
\label{scat}
\end{eqnarray}
in which $I_{if}(q_z)=\langle \psi_{zi}|e^{iq_{z} z}|\psi_{zf}\rangle$ and $q=|
E_{i}-E_{f}|/(\hbar v_{\lambda} \alpha)$.
$C_{LD}=\Xi^{2}\alpha^{3}/(8\pi\hbar v_{sl}^{2}D)$,
$C_{LP}=9e^{2}e_{14}^{2} \alpha \pi/(\hbar \kappa^{2}Dv_{sl}^{2})$ and
$C_{TP}=\pi e^{2}e_{14}^{2}\alpha/(\hbar\kappa^{2}Dv_{st}^{2})$
in Eq.\ (\ref{scat}) are the
coefficients from the electron-phonon scattering due to the deformation
potential and due to the piezoelectric coupling for the longitudinal
phonon mode and
two transverse phonon modes respectively.

In Fig.\ \ref{fig1} we plot the SRT of a typical double dot with $d_{0}=20$\ nm,
$a=10$\ nm, $d=5$\ nm, $V_{0}=0.4$ eV and $B=0.1$\ T at $T=4$\ K
as a function of electric field $E$.
The solid curve in Fig.\ \ref{fig1}(a)
is the result from the perturbation approach. It is interesting to see that
the SRT  is increased about {\em seven orders of magnitude} when the electric field is tuned
from $0.1$\ kV/cm to $1.3$\ kV/cm. The physics of such gate-voltage-induced
dramatic change can be understood
as follows:
When the gate voltage is small, due to the large well
height  $V_0$ and/or
large inter-dot distance $a$, the electron wavefunction (along the
$z$-axis) of the lowest
subband of each well is mostly localized in that well  due to the
high  barrier between them and hence the difference of the lowest
  two energy levels is very small (about $10^{-4}$\ eV).
When a gate voltage is high enough, electron can tunnel through the barrier
and the wavefunctions in the two wells get large overlap.
Therefore the separation between the lowest two
  levels $\varepsilon_1$ and $\varepsilon_2$ increases. This can be seen from
Fig.\ \ref{fig1}(b) where the energies of the lowest two levels
  along the z-axis $\varepsilon_1$ and $\varepsilon_2$
 are plotted against electric field $E$.
From Eqs.\ (\ref{E1}-\ref{E4}) one can see that the first two levels ($E_1$ and
$E_2$) and the next two levels ($E_3$ and $E_4$) are mainly separated by
the energy along the $z$-axis, {\em i.e.}, $\varepsilon_1$ and
$\varepsilon_2$. Such an increase makes the electron-phonon scattering more
efficient when the energy difference $\varepsilon_2-\varepsilon_1$ is not too big.
Therefore, by applying the gate voltage,
one finds the SRT first decreases. Nevertheless, with the further
increase of the gate voltage, half of the lowest four
 levels are quickly removed from the spin relaxation channel and the SRT
is enhanced. As a result, there is a minimum of SRT with the gate voltage.
This can be seen from the same figure where
 the weighted scattering rates ($\bar \Gamma_{i \to f}= f_{i}\Gamma_{i \to f}$)
 between different levels are plotted
versus the  electric field.
The leading contribution to the total scattering rate comes from
$\bar\Gamma_{3\to2}$ at small field regime. When the electric field
increases from $0.5$\ kV/cm to $1.3$\ kV/cm,
$\bar\Gamma_{3\to2}$ decreases rapidly
due to the separation
of $\varepsilon_\lambda$ with the electric field but
$\bar\Gamma_{1\to2}$ keeps almost unchanged as both levels $E_1$ and $E_2$
correspond to the same lowest level $\varepsilon_1$ along the $z$-axis. Finally
for large field, $\bar\Gamma_{1\to2}$ defines the total scattering
rate.  It is further noted that  although we performed the average of the 
initial and the sum of the final states in calculating the SRT,
the leading contribution  comes from the
scattering from $E_{3}$ to $E_{2}$ at low electric field
and the scattering  from $E_{1}$ to $E_{2}$ at large one.

The large variation of $\bar\Gamma_{3 \to 2}$ around 1\ kV/cm
can be estimated as following:
As the electron-phonon scattering due to the
piezoelectric coupling of the two transverse phonon modes
is at least one order of magnitude larger than the other modes, we only
consider the third term in Eq.\ (\ref{scat}).
From our calculation,
$\varepsilon_{1}=(3.25\times 10^{-4}E/(\mbox{kV/cm})+0.15129)$\ eV
and $\varepsilon_{2}=(1.68
\times 10^{-3}E/(\mbox{kV/cm})+ 0.1513)$\ eV. The energy splitting
between $E_{2}$ and $E_{3}$ can be approximated by
$\varepsilon_{2}-\varepsilon_{1}$. Therefore
$\Delta E_{23}=(1.36 \times
10^{-3} E/(\mbox{kV/cm})+5\times 10^{-5})$\ eV
 approximately and $q=\Delta E_{23}/(\hbar v_{st}
\alpha)$. As the variation of $|I_{12}(q\cos\theta)|$ in Eq.\ (\ref{scat})
is within one order of magnitude, we approximately  bring it
out of the integral. Then the remaining integral
$\int_{0}^{\pi/2}
d\theta \sin^{5}\theta (\sin^{4}\theta+8\cos^{4}\theta) e^{-q^{2}
  \sin^{2}\theta/2} $ can be carried out analytically:
$\frac{1}{2} B(\frac{1}{2};5)
 \Phi(5;\frac{11}{2}; -q^{2}/2)+4B(\frac{5}{2},3)\Phi(3;\frac{11}{2};
-q^{2}/2)$ with $B(\mu;\nu)$  and $\Phi(\alpha;
\gamma; z)$  being  the Beta function and the
degenerate Hypergeometric function separately.
When $E=0.1$\ kV/cm, the value of the integral is $10^{-1}$
and when $E=1.3$\ kV/cm, it becomes  $10^{-6}$.
Meanwhile, with the change of the electric field from
0.1\ kV/cm to 1.3\ kV/cm,
although $q^{3}|{\cal A}_{f}-{\cal B}_{i}|^{2}$ is increased by
one order of magnitude,
$|I_{23}|^{2}$ is decreased by one
order of magnitude and the distribution function $f_{3}$ is
decreased by another two
orders of magnitude. Therefore, $\bar\Gamma_{3\to2}$ decreases about seven
orders of magnitude when $E$ is tuned from 0.1\ kV/cm to 1.3\
kV/cm.

As pointed out by Cheng {\em et al.} \cite{Cheng} and confirmed by
Destefani and  Ulloa \cite{sergio}  that due to the strong spin-orbit
coupling, the perturbation  approach is inadequate in describing the SRT even
when the second-order energy corrections are included.
Therefore, in Fig.\ \ref{fig1}(a) we further
 plot the SRT calculated from the exact
diagonalization as dotted curve. Similar results are obtained although
again the SRT from the exact diagonalization approach differs from the
perturbation one.

\begin{figure}[htb]
\centerline
  {\psfig{figure=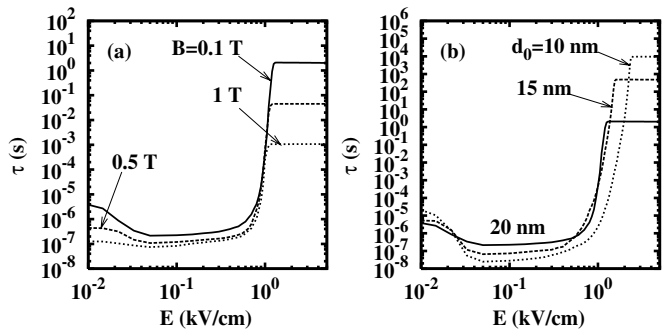,width=9cm,height=4.5cm,angle=0}}
\caption{SRT calculated from the exact diagonalization approach
  {\em vs.} the electric field at (a) different magnetic fields
    with $d_{0}=20$ nm
and (b) QD diameters with $B=0.1$ T.
In the calculation $a=10$\ nm, $d=5$\ nm, $d_{0}=20$ nm, $V_{0}=0.4$\
  eV and $T=4$\ K.}
\label{fig2}
\end{figure}

Now we investigate the magnetic field and dot size
dependence of the SRT in Fig.\ \ref{fig2}(a) and (b) by
exact diagonalization approach. Again one observes a dramatic
increase of the SRT by tuning the electric field up to a certain value
and then the SRT is insensitive to the electric field. For small
dot size ($d_0=10$\ nm), one even observes a {\em twelve orders of
magnitude} change of the SRT by tuning the gate electric field
to 2.6\ kV/cm. The dramatic variation of the SRT has been explained
above. Now we discuss why the SRT decreases with magnetic field and
dot size observed in Fig.\ \ref{fig2} in the electric-field-insensitive part.
From Fig.\ \ref{fig1}(b) one finds
 $\bar\Gamma_{1\to 2}$ is the leading contribution to the total
 scattering rate
in this part.
The energy splitting between the first and the second levels
$\Delta E_{12} \propto B$. As $\Delta E_{12}$ is about
$10^{-5}$ eV, $n_{q}\simeq k_{B}T/\Delta E_{12}$ and $n_{q} q^{3}
\propto (\Delta E_{12})^{2}$. Moreover $|{\cal A}_{1}- {\cal
  B}_{1}|^{2}= (\alpha \gamma_{1}^{\ast} 4 E_{B}\omega_{B})^{2}/(\hbar^{2}
\Omega \omega_{0}^{2})^{2}\propto B^{4}$ proximately. As a result, the
coefficient before the integral of the electron-transverse phonon scattering
due to the piezoelectric coupling is proportion to $B^{6}$.
Although the integral has a marginal decrease with $B$,
$\bar\Gamma_{1\to 2}$ still increases with $B$. Similarly, one can
explain the change of the SRT with the dot  diameter $d_{0}$.

\begin{figure}[htb]
  \centerline
  {\psfig{figure=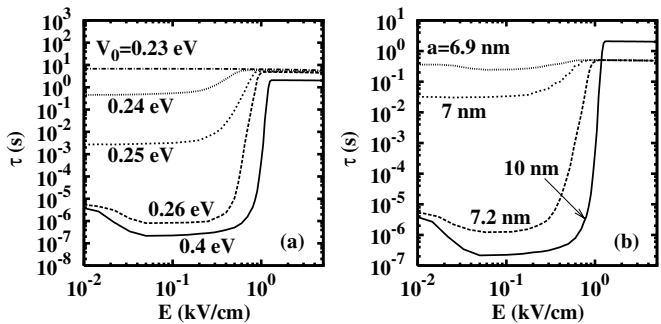,width=9.cm,height=4.5cm,angle=0}}
\caption{SRT calculated from the exact diagonalization
approach {\em vs.} the electric field at (a) different barrier heights $V_0$
with the barrier width $a=10$\ nm and (b) different barrier widthes $a$
with $V_0=0.4$\ nm.  In the calculation, $d=5$\ nm, $d_{0}=20$\ nm and $B=0.1$\ T.
 $T=4$\ K.
}
\label{fig3}
\end{figure}

It is noted that  in order to obtain the large variation of the SRT
by a gate voltage, it is important that the barrier between the
QDs should be large enough so that without a gate voltage, the
two dots are decoupled (and there is no energy
splitting along the $z$-axis). This can be clearly seen from Fig.\ \ref{fig3}:
With the decrease of the barrier height $V_0$ or the inter-dot distance $a$,
the tunability of the SRT by the gate voltage decreases.

\begin{figure}[htb]
  \centerline
  {\psfig{figure=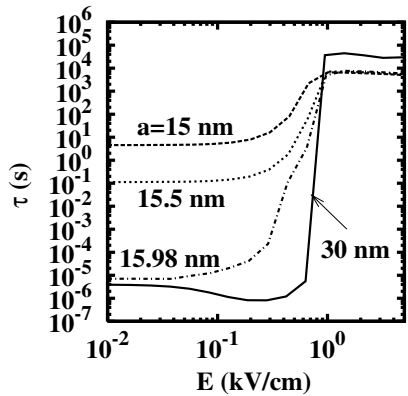,width=6cm,angle=0}}
\caption{SRT calculated from the exact diagonalization
approach {\em vs.} the electric field at different inter-dot distance
$a$ with low barrier height  $V_{0}=0.05$\ eV. In the calculation,
$d=5$\ nm, $d_{0}=20$\ nm, $B=0.1$\ T and $T=4$\ K.
}
\label{fig4}
\end{figure}

The double dot system proposed in our scheme can be easily realized
with the current technology.\cite{Hatano, Austing} Nevertheless, it is not
essential to use such a high barrier height system to obtain the
large spin manipulation. For ordinary barrier height widely used in the
experiment (which is about one order of magnitude lower than
$V_0$ used above), one can still achieve the similar manipulation
by increasing the distance $a$ between the two QDs as shown in Fig.\ 4
where the barrier height $V_{0}=0.05$\ eV.
One finds that for  small $V_{0}$,
if the barrier width $d$ is large enough, one can still get the large
change of SRT. Especially, in the case of $a=30$\ nm,
{\em eleven} orders of magnitude change of SRT is obtained
by a small gate field.

In conclusion, we have proposed a feasible scheme to manipulate the
spin relaxation in GaAs vertical double DQs by a small gate voltage.
The SRT calculated can be tuned up to twelve orders of
magnitude by an electric field from the gate voltage
less than 3\ kV/cm. This provides a unique way to control the
spin relaxation and to make spin-based logical gates. The
conditions to realize such a large tunability are addressed.
The double dot system proposed in our scheme can be easily realized
in the experiment. Finally the
proposed large orders of magnitude change due to the gate voltage
will not be reduced by the hyperfine
interaction with nuclear spins \cite{Erlingsson, Abalmassov}
as the SRT due to this mechanism in our case is around $10^3$\ s at 0.1\ T.
Finally we point out that differing from the earlier 
reports\cite{Denis,Tamborenea} where a strong variation of the SRT is
obtained from the anticrossing of the energy levels induced by the Rashba
spin-orbit coupling by increasing the magnetic field\cite{Denis} or 
the inter-dot distance,\cite{Tamborenea} 
there is no anticrossing/crossing of the 
energy levels in our scheme. 
Moreover, the tunability of the scheme proposed in the present paper is 
better as one only need to tune a very small gate voltage
(to tune the electric field from $0.1$ to $1.2$ kV/cm)
to obtain a surge of the SRT up to twelve orders of magnitude
in contrast to the large magnetic field of several tesla
to obtain the variation up to seven orders of magnitude.\cite{Denis}

This work was supported by the Natural Science Foundation of China
under Grant Nos. 90303012 and 10574120, the Natural Science Foundation
of Anhui Province under Grant No. 050460203, the Knowledge Innovation
Project of Chinese Academy of Sciences and SRFDP. The
authors would like to thank valuable discussions
with J. L. Cheng, J. Fabian, and X. D. Hu.


\begin{thebibliography}{0}
\bibitem{comment} It is noted that our final results
do not dependent on the choice of the confining potential.
 We have also considered other confining
  potential along the $z$-axis (for example, the double harmonic
  oscillator potential) and the results in this paper are still valid.
\end{thebibliography}
\end{document}